\documentstyle[12pt,aasms4]{article}
\begin{document}
\def \m {$ M_\odot$}
\def \l {$ L_\odot$}
\def \ro {g/cm$^{3}$}
\def \etal {\it et al.\rm}

\title{ CNO "Breakout" and Nucleosynthesis in Classical Novae } 

\author{S. Ami Glasner}
\affil{Racah Institute of Physics, The Hebrew University, Jerusalem,  Israel}

\author{ James W. Truran}
\affil{Department of Astronomy and Astrophysics, 
Enrico Fermi Institute, University of Chicago, Chicago, IL 60637, USA, 
and Physics Division, Argonne National Laboratory, Argonne, IL 60439, USA}

\author{Accepted for publication by the ApJL}
\begin{abstract}

For very slow white dwarf accretors in CV's Townsley and Bildsten (2004)
found a relation between the accretion rate
$\dot{M}$ and the central temperature of the white dwarf T$_c$. 
According to this relation for $\dot{M}$ less than $10^{-10} M_{\odot} yr^{-1}$
T$_c$ is much lower than $10^{7}$ K.
 Motivated by this study we follow the thermonuclear runaway on 
massive white dwarfs  
($M_{WD}=1.25 - 1.40$ \m) 
with T$_c$ lower than $10^{7}$ K, accreting matter of solar composition.
We demonstrate that in that range of the relevant parameter space  
($Tc,M_{WD}$ and $\dot{M}$) the 
slope of the relation between the
peak temperatures achieved during the runaway 
and T$_c$ becomes much steeper than its value for T$_c$ above $10^{7}$ K.
The peak temperatures we derive can lead to 
nuclear breakout from the conventional 
"hot carbon-nitrogen-oxygen" cycle. When breakout conditions are achieved 
the heavy element abundances can show a much wider variety than what
is possible with the common enrichment mechanisms.

\end{abstract}


\section {Introduction}
 
Classical nova explosions are a consequence of the unstable ignition of 
hydrogen in accreted envelopes on white dwarfs in close binary systems. 
Peak temperatures in the range $\approx 200-350$ million K 
achieved in the ensuing 
thermonuclear runaway yield burning of hydrogen to helium by means of the 
hot CNO cycles, with the associated nuclear energy 
release $\approx 5 \times 10^{18}$ 
erg g$^{-1}$ powering the outburst. One dimensional numerical simulations 
\cite{SSt74,SSt85,Macd80,Macd83,nar80,pss78,pri86,ibn82,tru82,JorHer98}
have established that many features of classical novae in outburst can be 
understood on this model, for representative choices of white dwarf mass 
($M_{WD} \approx  M_\odot$) 
and accretion rate 
($ \dot{M} \approx 10^{-10}-10^{-9} M_\odot yr^{-1} $ ). 
More details can be found in a review by \cite{star02}.

Observational studies of novae reveal that, while the composition of the 
accreted matter is consistent with Solar abundances, the ejecta are 
characterized by high concentrations of heavy 
elements (e.g. C, N, O, Ne, Mg) relative to solar abundances \cite{gtw98}.
For most nova models, numerical simulations predict that peak temperatures 
achieved in the runaways do not exceed the 
critical value (T$_{crit}$ of $\approx $
4 $\times$ 10$^8$ K) at which "breakout" from the hot CNO cycles can occur 
\cite{wiescher86,YaPrial05}. This has led to the conclusion accepted 
by most nova theorists that the observed enhancements reflect the effects 
of "dredge-up" of CO- or ONeMg-rich matter from the underlying white 
dwarf star \cite{LivTru90}. We show here that in a restricted, and therefore rare,
region of the relevant parameter space ( $Tc,M_{WD}$ and $\dot{M}$ ) breakout burning 
can lead to heavy element enrichment caused only by breakout burning.

In a recent exploration of the thermal state of accreting white dwarfs 
experiencing classical nova explosions, Townsley and Bildsten (2004) 
have demonstrated that there exists a population of massive white dwarfs  
($M > 1.2 M_\odot$) accreting at rates as 
low as $\approx 10^{-11} M_{\odot} yr^{-1}$, with central temperatures 
$T_c$ of $\approx 5 \times 10^{6}$ K. The combination of cold  
white dwarfs with low accretion rates
leads to novae runaways with massive envelopes and exceptionally high
peak temperatures. We show here that for a significant part of the
relevant parameter space ($Tc,M_{WD}$ and $\dot{M}$) the peak temperature
at runaway, T$_{peak}$ is well above the minimum value for CNO ``Breakout''.
 
In section II we briefly describe the hydro solver and the reaction network.
Section III presents the main results of our models. 
We conclude with a discussion of the possible observational consequence 
of our results and of their consistency with previously published models.

\section  {The hydrodynamic solver}

The accretion process and the runaway are studied with a 1D Lagrangian 
implicit code that integrates the equations of momentum and energy 
conservation assuming spherical symmetry. The energy transfer  includes 
a radiative term and a convective term. The radiative component is
a diffusive flux. The radiative diffusion coefficient is determined 
according to  the Iglasias \& Rogers opacity tables  \cite{IGR96} for 
temperatures above $Log(T)=3.75$ Kelvin and according to the Alexander fit 
for lower temperatures. Electron conductive opacities are computed according to the Itoh fit \cite{Itoh83}. 
The convective energy flux is computed 
according to the mixing length theory \cite{spi63}. Within a convective 
zone the matter is mixed using a diffusion coefficient that takes into 
account both the local convective velocity and the mixing length 
(which is taken to be two pressure scale heights): 

$$
D_{c} = v_{c}l/3
$$  

where $D_{c}$ is the diffusion coefficient, 
$v_{c}$ is the 1D convective velocity, derived from the mixing 
length theory and $l$ is the effective mixing length. 

In the present survey we ignore element settlement by diffusion and 
undershoot mixing. The equation of state employed to derive the energy 
and pressure for given temperature density and chemical abundances is 
an ideal gas for the ions and involves a table fit for the numerical 
integration of the Fermi integrals for the electrons (and positrons).
The rate of nuclear energy production and the abundance changes are 
calculated with a 160 nuclei network. The proton-proton reaction rate 
in the PeP channel take into account the extra electron density term 
due to the electron capture \cite{Bahcal69}. The chemical elements 
included in the network were chosen in a proper way to provide an 
accurate history of the proton-rich composition of the accreted envelope. 
 
Accreted material is added each time step to the outermost zone as 
dictated by the accretion rate and the size of the time step. The 
outermost zone in the numerical grid  is divided into two grid zones 
whenever the mass of the zone is greater than a specified value 
(taken to be twice the specified mass resolution in the envelope). 
The mass is added with the instantaneous thermodynamic properties of 
the outermost grid zone. The same 1D solver was used in order to produce
the initial models for all our 2D novae studies \cite{gl95,glt97,glt07}.
More details about the solver and its abilities will be given in a 
forthcoming paper.  

\section  {The main results}

In this study we follow the accretion process and the ensuing runaway 
for cold massive white dwarfs accreting at very low rates. Under such 
conditions the peak temperature achieved in the runaway is expected 
to be quite high. In order to trace the nucleosynthesis history, we 
solve the hydrodynamic equations coupled with a nuclear reaction 
network that includes 160 nuclei up to $^{64}Ni$. As we noted in the 
introduction, our interest in this poorly explored region of the nova 
parameter space was motivated by the results published by \cite{TB2004}.
In our survey, we include white dwarf cores in the mass  range 
1.2 $ M_\odot$ to 1.4 $M_\odot$  accreting matter of solar composition 
\cite{AG89} at rates ranging  from  $10^{-11} M_\odot yr^{-1}$ 
to $10^{-10} M_\odot yr^{-1}$. The initial models are cold white dwarf 
cores in hydrostatic and thermal equilibrium characterized by their 
central temperature (T$_c$) with values ranging 
from $ T_c=4 \times 10^6$ K  to $ T_c=2 \times 10^7$ K.
 
The combination of very low central temperatures T$_c$ and very slow 
accretion rates, which demands the use of a greatly expanded nuclear
reaction network, is the unique feature of this survey. In this short 
initial report we therefore focus on the main outcome: the extremely 
high peak temperatures achieved in the runaway and the induced 
nucleosynthesis. The main features of the time history of the total 
burning rate, the bolometric luminosity, and the peak temperature 
at the base of the envelope are quite similar for all studies models. 

Prior to the runaway, some of the considered models accrete exceptionally 
high envelope masses $\delta M_{ignite}$. The ensuing thermonuclear 
runaways are characterized by very high peak temperatures, $T_{peak}$. 
In Figure 1 we present the relation between $T_{peak}$ and T$_c$ for 
white dwarf masses and accretion rates ($\dot{M}$) that are relevant 
to this work; in Figure 2 we show the dependence of $\delta M_{ignite}$ on 
the same variables. The common feature for all the considered models is 
the increase of $T_{peak}$ with decreasing T$_c$:  $T_{peak}$ is greater 
than T$_{crit}$ only for very cool white dwarfs ($T_c \le 5-7 \times 10^{6}$).
 
The high temperature and high density at the bottom of the accreted 
envelope and the comparatively long timescale for which these conditions 
prevail, defined as  $\delta t_{break}$, allow a significant epoch 
of explosive hydrogen burning. In Table I we present the main input 
parameters of our models - $M_{WD}$, T$_c$, and $\dot{M}$ -  together  
with the major hydrodynamic outcome quantities that are relevant to the 
burning: $\delta M_{ignite}$ , $T_{peak}$ and  $\delta t_{break}$ .
 
As expected, hydrogen burning under such extreme conditions yields 
``breakout'' from the conventional ``hot carbon-nitrogen-oxygen" cycle.
During most of the time that the temperature stays above T$_{crit}$ 
the envelope is fully convective. Therefore, the nuclei produced in the 
burning are spread homogeneously over the entire envelope. Once the shell 
temperature falls below T$_{crit}$ the overall abundance of heavy nuclei 
is not expected to change during the later stages of the runaway. 
We selected one representative model from Table I 
(s135u for M=1.35 \m ) and present its initial 
and final abundance patterns as a function of mass number (A) in Figure 3. 
We note that both the mass of heavy elements ejected and their 
distribution depend on the details of the mass ejection mechanism; we 
will present a study of the ejection phase in a forthcoming paper.

 \section  {Discussion}

There is a well established consensus in the nova research community 
that the relevant parameters that define the thermonuclear runaway 
are the white dwarf mass ($M_{WD}$), the central temperature (T$_c$), 
and the accretion rate ($\dot{M}$), together with the composition of both 
the cold white dwarf and the accreted matter. This parameter space has 
been well explored by many researchers. Most of the previous studies 
were concerned with the range of central temperature above approximately 
T$_c$ = 10$^7$ K.
But, in a recent study, Townsley 
and Bildsten (2004) demonstrate that for very slow accretors there exists 
a steady state that determines a relation between $\dot{M}$ and T$_c$. 
For $\dot{M}$ less than $10^{-10} M_{\odot} yr^{-1}$, they find that 
T$_c$ according is much smaller than $10^{7}$ K. We therefore focused 
our study on massive white dwarfs with T$_c$ much below T$_c$=10$^7$ K 
and very slow accretion rates.
 
Since the expected temperatures at the runaway are above T$_{crit}$, 
the nuclear energetics and associated nucleosynthesis in our models 
were calculated with the use of an extended nuclear reaction network 
that included 160 nuclei up to $^{64}Ni$. Our results are generally 
consistent with those of the 1.3 \m model of Nariai et al. (1980), as
is evident from our Table I. For T$_c$=10$^7$ K, both the accreted mass 
we obtain and the peak temperature $T_{peak}$ achieved during the 
runaway are consistent with the 1.35 \m model of Jose and Hernanz (1998) 
and with the 1.25 \m and 1.40 \m models of Yaron et al. (2005). For the 
conditions of central temperature and accretion rate we took from the 
results of Townsley and Bildsten (2004), the ignition mass we obtained 
is fully consistent with their stated results 
(compare our model s125y in Table I with their Figure 8).
 
Considering the dynamical aspects of our survey, we find a very steep 
dependence of the accreted mass and peak temperature at runaway 
on the central temperature T$_c$ below T$_c$=10$^7$ K. In contrast, 
the dependence in the more commonly studied region above T$_c$=10$^7$ K
is very shallow. This behavior can be understood on the basis of the well 
known dependence of the burning rates on the temperature during the 
accretion stage. For the accreted matter of assumed solar 
composition, the slope of the logarithmic derivative of the burning rate 
changes in the vicinity of a temperature ~ 10$^{7}$ from 4 (PP cycle) 
to about 18 (CNO cycle). At central temperatures $T_c=4 \times 10^6$ K,
relevant to cataclysmic variables below the period gap \cite{TB2004}, 
the accreted envelope mass required to initiate runaway can be 4 to 5 
times  greater than for $T_c=10^7$ K (Table I). This result 
confirms that once the core of the white dwarf is sufficiently cool,
massive hydrogen-rich accreted envelopes can exist at runaway even 
for very massive white dwarfs.  Therefore, observational evidence for 
nova outbursts with massive accreted envelopes in the range of a few 
times $10^{-4}$ \m cannot exclude very massive white dwarfs as the 
underlying core.
 
 Most significant are the results we obtain for the nucleosynthesis 
occurring in our models. The high temperatures, much above  T$_{crit}$, 
that can prevail for thousands of seconds lead to a very interesting 
path of nuclear burning up to the iron peak region. In order to identify 
the pattern of enrichment of iron group elements, we divided the elements 
included in our models into five groups: (1) Hydrogen; (2) Helium;
(3) all the CNO cycle elements up to fluorine;  (4) the elements from neon 
to calcium (the intermediate mass elements); and (5) the iron group elements 
from titanium to nickel. For all the models reported in Table I, the 
peak temperature in the runaway exceeds T$_{crit}$ and produces large 
concentrations of iron group elements. The concentrations of freshly 
synthesized iron group elements (relative to the original solar amounts) 
range from an increase by a factor of 1.75 for the less massive models 
(model s125y with M=1.25 \m) to an increase by a factor of 10.0 for the 
more massive models (model s135y with M=1.35 \m); this can be seen from 
the results presented in table II. Both CNO elements 
(from breakout) and intermediate mass elements are seen to be transformed 
to iron group elements. The outcome is a net decrease in the mass 
of hydrogen and CNO elements.  

As one might expect, for the higher mass cases (higher peak temperature) 
there is even  a small decrease in the fraction of intermediate mass 
elements and a significant production of iron group elements (see e.g. 
Figure 3 and Table II). For lower mass white dwarfs (lower peak 
temperature) there is a slight increase in the mass fraction of 
intermediate mass elements and a smaller level of production of iron group 
elements (see e.g. Table II).

In order to make certain that we indeed achieved ``breakout'' from the 
conventional "hot carbon-nitrogen-oxygen" cycle we made a definite test.
A model similar to s135u  (M=1.35 \m), for which the accreted matter 
was assumed to include solar concentrations of nuclei only up to fluorine, 
was evolved. The energetics of this model are similar to those of model
s135u (Table I). With regard to the nucleosynthesis, the final 
concentration of iron group elements was determined to be comparable 
to that of model s135u. There is only a small concentration of intermediate 
mass elements in the final stages, which confirms that indeed CNO nuclei 
were burned all the way to iron.  

Our main conclusion concerning nucleosynthesis is that due to  ``breakout'' 
from the conventional "hot carbon-nitrogen-oxygen" cycle, the heavy element
abundances in the relevant novae ejecta can show a much wider variety
than what is possible with the conventional mixing mechanisms. Some rare
cases could be enriched by breakout alone.  
Future work will therefore concentrate on the observational predictions 
that can be made about the time history of the ejected mass. In order to test this
hypothesis we will also consider 
the interpretation of observed abundance patterns in such nova systems as 
Nova Cygni 1992 and Nova Her 1991, in light of the possibility that they 
may well have experienced nuclear burning at temperatures approximating 
or above breakout leading to increased nucleosynthesis of intermediate 
mass and heavy elements.  In order to improve the accuracy of such 
predictions, we intend to follow the abundance history of the ejected 
matter all through the phase of nuclear burning and mass ejection.

\clearpage
\centerline{Acknowledgments}

This work is supported in part at the University of Chicago by
the Department of Energy under Grant B523820 to the ASC/Alliances
Center for Astrophysical Thermonuclear Flashes and by the National
Science Foundation under Grant PHY 02-16783 for the Frontier Center
"Joint Institute for Nuclear Astrophysics" (JINA), and at the Argonne
National Laboratory by the U.S. Department of Energy, Office of
Nuclear Physics, under contract DE-AC02-06CH11357.

\begin{deluxetable}{crrrrrr}
\footnotesize
\tablecaption{The main parameters for all models that pass T$_{crit}$ \label{tbl-1}}
\tablewidth{0pt}
\tablehead{
 \colhead [MODEL] & 
\colhead{$M_{WD}$ [\m]}  &   \colhead{Tc [$10^6$ K]}  & \colhead{$\dot{M}$ [\m yr$^{-1}$]}  & \colhead{$\delta M_{ignite}$ [\m]}  & \colhead{$T_{peak}$ [$10^8$ K]}
& \colhead{$\delta t_{break}$ [sec] }
 }
\startdata
 s135u & $ 1.35$ & $  4.00 $ &  $  10^{-11}  $   &  $ 6.38\times 10^{-5}  $   &  $ 5.14 $      & $ 14500. $  \nl
 m135k & $ 1.35$ & $  5.88 $ &  $  10^{-11}  $   &  $ 3.90\times 10^{-5}  $   &  $ 4.57 $      & $ 12000. $  \nl
 m135j & $ 1.35$ & $  7.65 $ &  $  10^{-11}  $   &  $ 2.70\times 10^{-5}  $   &  $ 4.23 $      & $  9500. $  \nl
 m135i & $ 1.35$ & $  9.03 $ &  $  10^{-11}  $   &  $ 2.06\times 10^{-5}  $   &  $ 4.00 $      & $  7000. $  \nl
 m135r & $ 1.35$ & $  4.00 $ &  $  10^{-10}  $   &  $ 3.40\times 10^{-5}  $   &  $ 4.42 $      & $ 11000. $  \nl
 m135q & $ 1.35$ & $  5.88 $ &  $  10^{-10}  $   &  $ 2.63\times 10^{-5}  $   &  $ 4.16 $      & $  9000. $  \nl
 m135p & $ 1.35$ & $  7.65 $ &  $  10^{-10}  $   &  $ 2.01\times 10^{-5}  $   &  $ 3.92 $      & $  7000. $  \nl
 m135o & $ 1.35$ & $  9.03 $ &  $  10^{-10}  $   &  $ 1.61\times 10^{-5}  $   &  $ 3.51 $      & $  1000. $  \nl
 m13ek & $ 1.30$ & $  4.00 $ &  $  10^{-11}  $   &  $ 9.62\times 10^{-5}  $   &  $ 4.29 $      &  $ 7000. $  \nl
 m13ed & $ 1.30$ & $  6.00 $ &  $  10^{-11}  $   &  $ 5.41\times 10^{-5}  $   &  $ 3.82 $      &  $ 4250. $  \nl
 s125y & $ 1.25$ & $  4.00 $ &  $  10^{-11}  $   &  $ 13.02\times 10^{-5} $   &  $ 3.68 $      &  $ 3000. $  \nl

\enddata
\end{deluxetable}

\begin{deluxetable}{ccrr}
\footnotesize
\tablecaption{The changes in  mass abundances grouped according to mass number for the models
 s135u and s125y \label{tbl-2}}
\tablewidth{0pt}
\tablehead{
 \colhead{jobname} & \colhead{group of elements} & \colhead{absolute change} & \colhead{relative change}
 }
 \startdata
  s135u  & $\Delta X[H]     $  & $ -0.0740 $ & $ -0.10  $ \nl
    -    & $\Delta X[He]    $  & $ +0.0660 $ & $ +0.24  $ \nl
    -    & $\Delta X[C-F]   $  & $ -0.0023 $ & $ -0.23  $ \nl
    -    & $\Delta X[Ne-Ca] $  & $ -0.0032 $ & $ -0.90  $ \nl
    -    & $\Delta X[Ti-Ni] $  & $ +0.0135 $ & $ +10.0  $ \nl
  s125y  & $\Delta X[H]     $  & $ -0.0377 $ & $ -0.05  $ \nl
    -    & $\Delta X[He]    $  & $ +0.0365 $ & $ +0.13  $ \nl
    -    & $\Delta X[C-F]   $  & $ -0.0007 $ & $ -0.07  $ \nl
    -    & $\Delta X[Ne-Ca] $  & $ +0.0009 $ & $ +0.25  $ \nl
    -    & $\Delta X[Ti-Ni] $  & $ +0.0010 $ & $ +0.75  $ \nl
\enddata

\end{deluxetable}

\clearpage
\newpage
\plotone{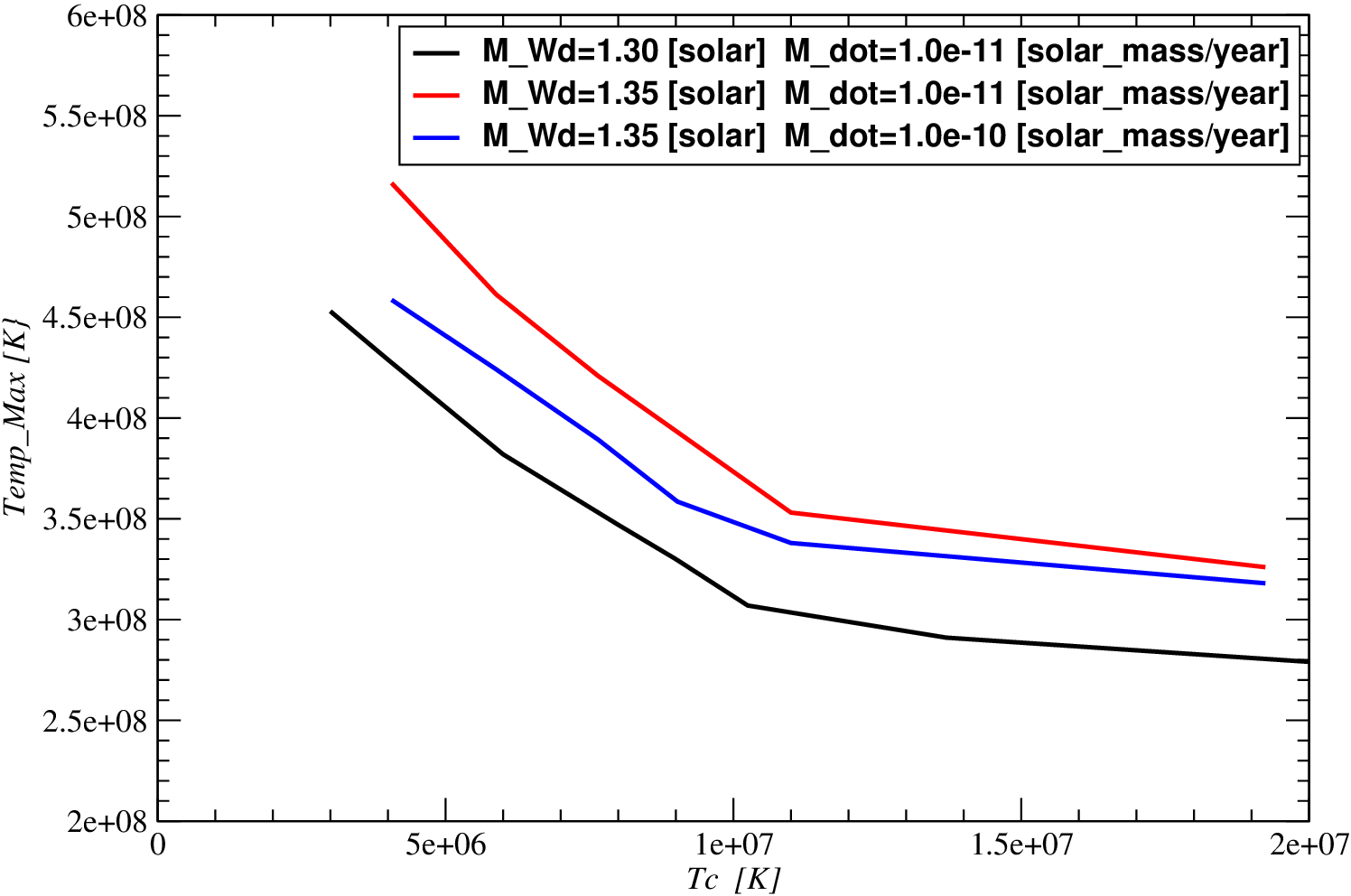}
 \figcaption[f1.eps]{The peak temperature achieved in the runaway
as a function of the central temperature T$_c$ of the white dwarf.
 \label{fig1}}
\newpage
\plotone{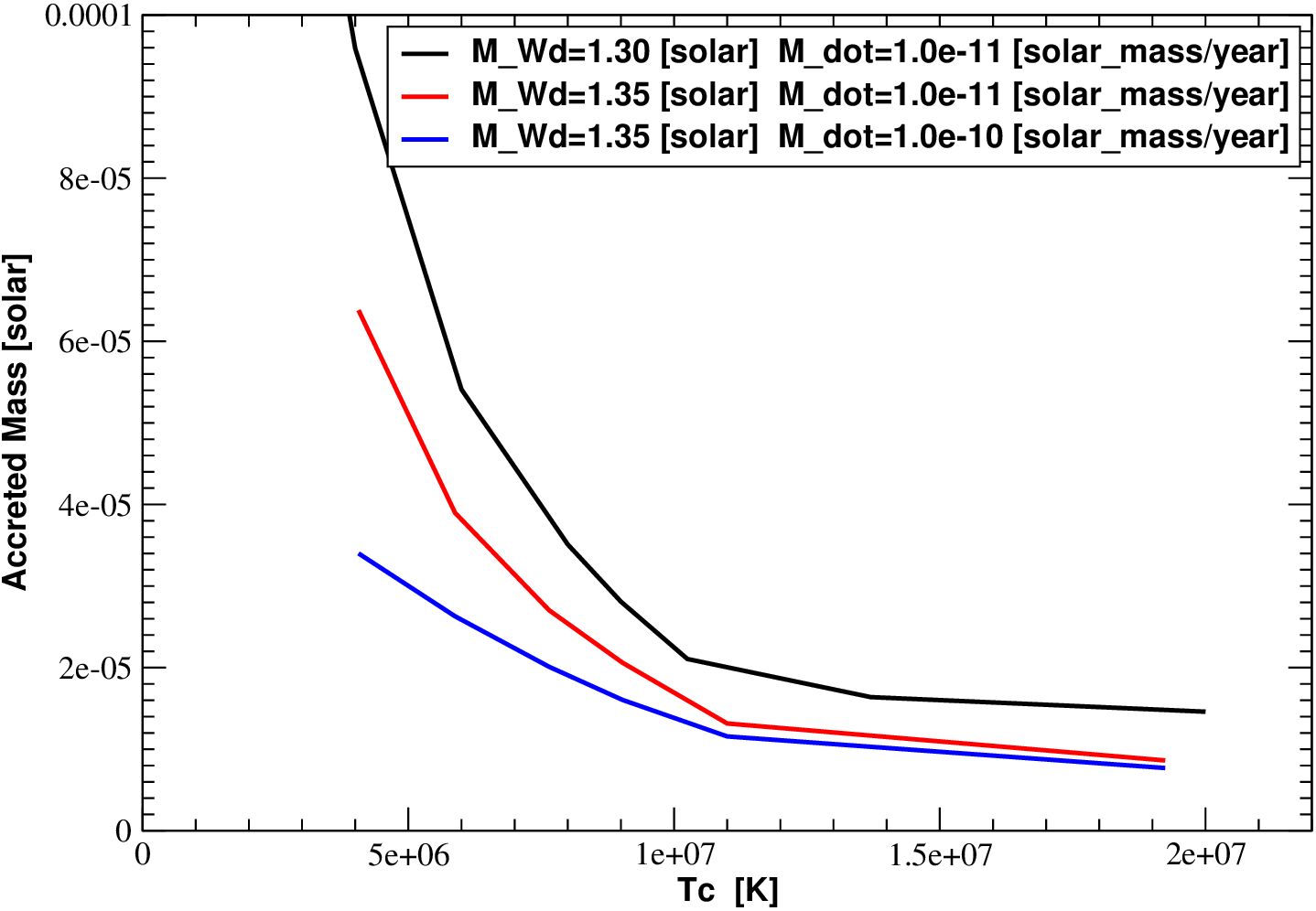}
 \figcaption[f2.eps]{The accreted mass at ignition
as a function of the central temperature T$_c$ of the white dwarf.
 \label{fig2}}
\newpage
\plotone{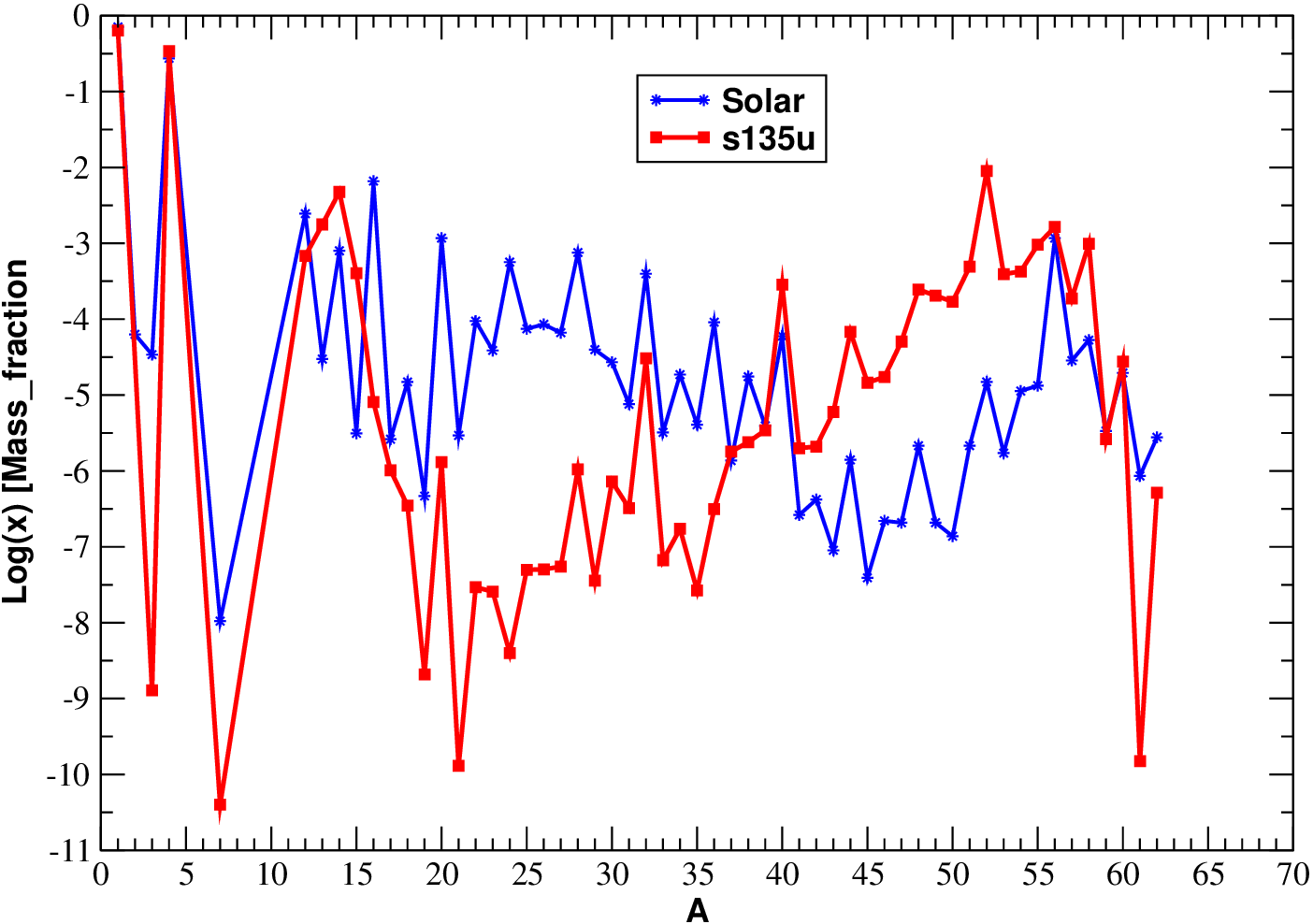}
 \figcaption[f2.eps]{The abundances of the elements in the reaction network  
as a function of the mass number A. Presented are both the initial 
(solar) abundances and the `final' abundances when the peak temperature 
has fallen below T$_{crit}$ (Model s135u, M=1.35 \m).
 \label{fig3}}
\newpage
\end{document}